\documentclass[a4paper,fleqn,letters,usenatbib]{mnras}

\usepackage{newtxtext,newtxmath}
\usepackage{mathptmx}
\usepackage{txfonts}
\usepackage[T1]{fontenc}
\usepackage{ae,aecompl}
\usepackage{graphicx}	
\usepackage{amsmath}	
\usepackage{amssymb}	

\newcommand{\vrel}{V_\mathrm{rel}}
\newcommand{\vfrag}{V_\mathrm{frag}}

\title{ALMA images of discs: are all gaps carved by planets?}

\author[J.-F. Gonzalez et al.]{
J.-F. Gonzalez,$^{1}$\thanks{E-mail: jean-francois.gonzalez@ens-lyon.fr}
G. Laibe,$^{2}$
S. T. Maddison,$^{3}$
C. Pinte$^{4,5}$
and F. M\'enard$^{4,5}$
\\
$^{1}$Universit\'e de Lyon, Lyon, F-69003, France ; Universit\'e Lyon~1, Observatoire de Lyon, 9 avenue Charles Andr\'e, Saint-Genis Laval, F-69230,\\ France ; CNRS, UMR 5574, Centre de Recherche Astrophysique
de Lyon ; \'Ecole Normale Sup\'erieure de Lyon, Lyon, F-69007, France\\
$^{2}$School of Physics and Astronomy, University of Saint Andrews,
North Haugh, St Andrews, Fife KY16 9SS, United Kingdom\\
$^{3}$Centre for Astrophysics and Supercomputing, Swinburne University of Technology, PO Box 218, Hawthorn, VIC 3122, Australia\\
$^{4}$UMI-FCA, CNRS/INSU France (UMI 3386), and Departamento de Astronom{\'\i}a, Universidad de Chile, Casilla 36-D Santiago, Chile\\
$^{5}$UJF-Grenoble 1 / CNRS-INSU, Institut de Plan\'etologie et d'Astrophysique de Grenoble, UMR 5274, Grenoble, F-38041, France
}

\date{Accepted 2015 August 12. Received 2015 August 05; in original form 2015 July 15}

\pubyear{2015}

\begin{document}
\label{firstpage}
\pagerange{\pageref{firstpage}--\pageref{lastpage}}
\maketitle

\begin{abstract}
Protoplanetary discs are now routinely observed and exoplanets, after the numerous indirect discoveries, are starting to be directly imaged. To better understand the planet formation process, the next step is the detection of forming planets or of signposts of young planets still in their disc, such as gaps. A spectacular example is the ALMA science verification image of HL Tau showing numerous gaps and rings in its disc.

To study the observability of planet gaps, we ran 3D hydrodynamical simulations of a gas and dust disc containing a 5~$M_\mathrm{J}$ gap-opening planet and characterised the spatial distribution of migrating, growing and fragmenting dust grains. We then computed the corresponding synthetic images for ALMA. For a value of the dust fragmentation threshold of 15~m\,s$^{-1}$ for the collisional velocity, we identify for the first time a self-induced dust pile up in simulations taking fragmentation into account. This feature, in addition to the easily detected planet gap, causes a second apparent gap that could be mistaken for the signature of a second planet. It is therefore essential to be cautious in the interpretation of gap detections.\end{abstract}




\begin{figure*}
\centering
\resizebox{\hsize}{!}{
\includegraphics{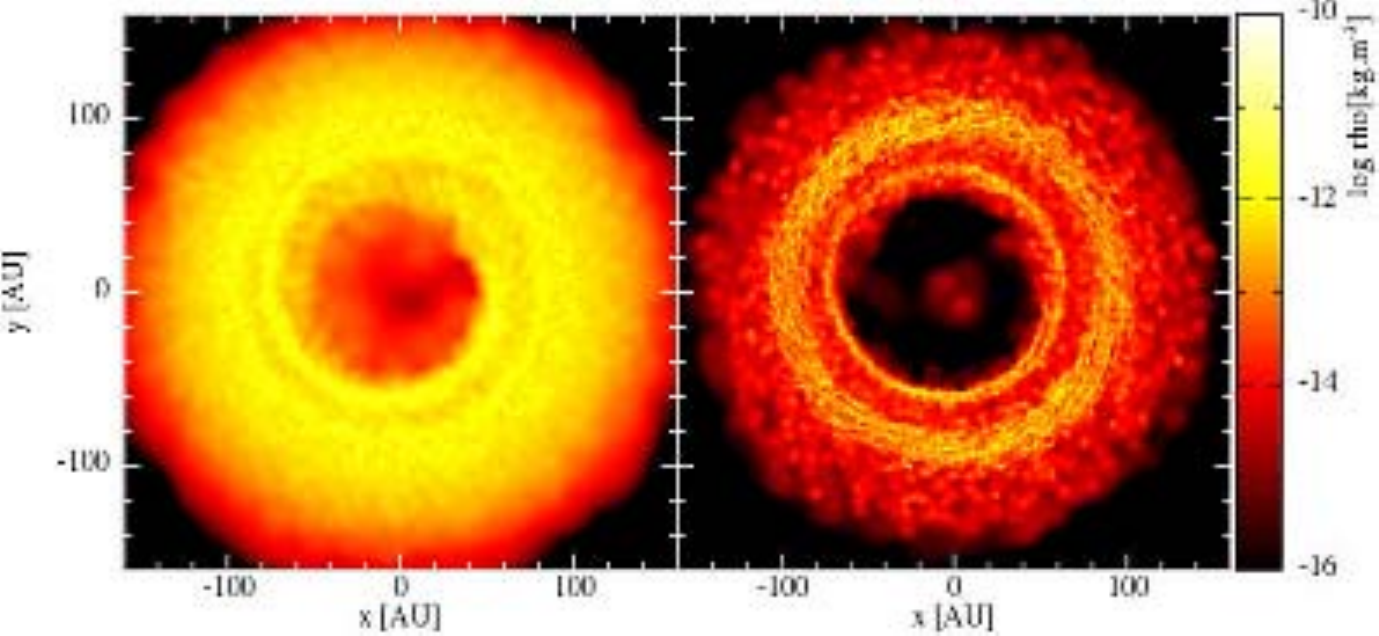}
\hspace{2em}
\includegraphics{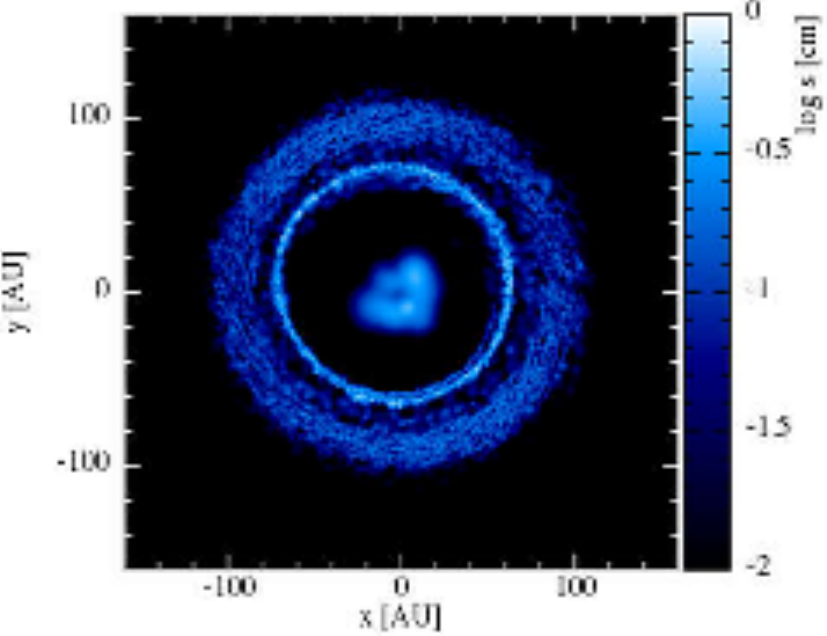}
}
\vspace{-4ex}
\caption{Maps of the volume density of the gas (left) and dust (centre) and of the average dust grain size (computed with the SPH kernel, right) in the disc midplane after 100\,000~yr.}
\label{Fig:hydro}
\end{figure*}

\section{Introduction}
\label{sec:Introduction}

The holy grail in the field of planetary science is to catch planet formation in action, and the most anticipated signpost of massive planets in discs are gaps. Many gas-rich protoplanetary discs have now been observed with gaps and inner cavities, especially at sub-millimetre and millimetre wavelengths \citep[e.g.][]{Casassus2013, Canovas2015,vanderMarel2015}. These so-called transitional discs have inner regions which have undergone substantial clearing, due to a combination of grain growth, photoevaporation and/or interaction with an embedded planet or companion \citep{Espaillat14}.  While there are now a few examples of discs which show tantalising evidence of embedded protoplanets and disc clearing (e.g. LkCa~15, \citealt{KrausIreland12}; HD~100546, \citealt{Quanz2015}) direct observational evidence of planet clearing remains elusive.  Planet gaps have been predicted theoretically for several decades \citep{PapLin84} and many numerical studies of planets opening gaps have been conducted \citep[e.g.][]{Kley99,Paardekooper2004,dVB06,Crida2006,Fouchet2007,Ayliffe12}, as have predictions of their observability \citep{Wolf2002,Gonzalez2012,Ruge2013}.

More recently, the spectacular ALMA science verification images of HL~Tau \citep{HLTau2015} surprised everyone with its clearly discerned multiple rings and gaps, which even more surprisingly are axisymmetric.  It is tempting to attribute each ring plus gap in the ALMA image of HL~Tau to a planet, but is this the right thing to do? \citet{Zhang2015} invoke potential accumulation of dust at condensation fronts. However, the location of the first three gaps observed by ALMA in HL Tau are found to be close to mean motion resonances, suggesting that the rings could be generated by planets \citep{HLTau2015}. The question we address in this Letter is whether all gaps seen in protoplanetary discs are always planet gaps, or could they result from alternative processes occurring during the early stages of planet formation? We explore here the effects of dust pile up as such an alternative.

Millimetre continuum images trace the spatial distribution of dust grains. Their evolution in discs is governed by two parameters. The first is the Stokes number, St, which is the ratio of the drag stopping time to the orbital period. The value of St determines the efficiency of vertical settling, radial drift and growth/fragmentation of grains in the disc \citep[e.g.][]{W77,NSH1986,SV1997,Youdin2002,Garaud2004,Nader2005,Brauer08,Birnstiel10,Laibe2014}. In particular, grain migration is fastest when $\mathrm{St}\sim1$. St is a function of the grain size and density, but depends also on the local gas conditions. As such, it varies as a function of the distance to the central star for a given grain size. HL Tau is a young and massive extended disc supplied by in-falling material from the surrounding envelope. For such discs, millimetre-sized grains have $\mathrm{St}\sim1$ at a few tens of au, at which point the dust and gas strongly decouple \citep{Laibe2012}. The second parameter involved is the dust-to-gas ratio\,$\epsilon$. For $\epsilon$ of order unity, dust affects the gas as much as gas affects dust due to aerodynamic drag. Although the dust backreaction due to dust inertia is negligible in most astrophysical systems, it becomes critical in regions of discs where dust concentrates, such as particles traps.

In this Letter we present a mechanism where grain growth and fragmentation result in dust distributions that produce apparent gaps when the dust backreaction is taken into consideration. We provide the details on our numerical simulations and synthetic observables in Sect.~\ref{sec:Methods}, discuss the origin of the apparent gap in Sect.~\ref{sec:Discussion}, and conclude in Sect.~\ref{sec:Conclusion}.

\section{Methods and results}
\label{sec:Methods}

\subsection{Hydrodynamic simulations}
\label{sec:Hydro}

We study the evolution of gas and dust in protoplanetary discs using our 3D, two-phase, Smoothed Particle Hydrodynamics (SPH) code, described in \citet{BF2005}. We self-consistently treat the aerodynamical drag between gas and dust, including the backreaction of dust on gas. We follow the standard SPH implementation of artificial viscosity \citep{Monaghan1989} and use $\alpha_\mathrm{\scriptscriptstyle SPH}=0.1$ and $\beta_\mathrm{\scriptscriptstyle SPH}=0.5$, corresponding to a uniform \citet{SS1973} parameter $\alpha_\mathrm{\scriptscriptstyle SS}\sim10^{-2}$ \citep[see][for a discussion]{Fouchet2007}. \citet{Arena2013} showed that the expected properties of turbulence are naturally reproduced by the SPH formalism. Grain growth is treated as described in \citet{Laibe2008}, via the prescription developed by \citet{SV1997} based on the determination of the relative velocity of grains, $\vrel\propto c_\mathrm{s}\sqrt{\alpha_\mathrm{\scriptscriptstyle SS}\mathrm{St}}/(1+\mathrm{St})$, as a function of local properties of the gas phase (e.g. its sound speed $c_\mathrm{s}$) and grain size. We implement grain fragmentation by defining a velocity threshold, $\vfrag$, to which we compare $\vrel$. Collisions occurring at speeds below the threshold velocity lead to grain growth, while collisions at greater speeds result in the shattering of grains, which we model by decreasing the size of the representative grains. The numerical procedure is detailed in \citet{Gonzalez2015}.

We model a disc typical of Classical T Tauri Stars (CTTS) of mass $M_\mathrm{disc}=0.01\ M_\odot$ around a 1~$M_\odot$ star, with an initially uniform dust-to-gas ratio $\epsilon=10^{-2}$. The disc contains a 5~$M_\mathrm{J}$ planet on a circular orbit at a fixed radius of $r_\mathrm{p}=40$~au. The planet is implemented as a point mass particle subject only to the star's gravity, whereas the SPH gas and dust particles feel the gravitational potential of the star and planet \citep[see][]{Fouchet2007}. The disc is initially set up as in \citet{Fouchet2010} with a constant surface density $\Sigma(r)=19.67$~kg\,m$^{-2}$ with radial extension from 4 to 120~au, a vertically isothermal temperature profile $T(r)=15\,(r/r_\mathrm{p})^{-1}$~K, and is free to evolve. Grains have an initially uniform size $s_0=10\ \mu$m and are allowed to grow and fragment. In  \citet{Gonzalez2015}, we presented a suite of simulations that vary the fragmentation threshold: $\vfrag=10$, 15, 20, 25~m\,s$^{-1}$ and $+\infty$ (i.e.\ growth only), evolved for 100\,000~yr or $\sim400$ planetary orbits, and discussed our choice of parameters and their influence on the results. In particular, our disc temperature is colder than what observations suggest \citep[e.g.][]{Williams2014}. Changing its value would affect gap formation (only moderately for a high-mass planet), as well as fragmentation ($\vrel\propto c_\mathrm{s}\propto\sqrt{T}$). In the midplane, St is independent of $T$, leaving dust dynamics unchanged. In this Letter, we focus on the $\vfrag=15$~m\,s$^{-1}$ case, for which we see a second apparent gap.

\begin{figure}
\centering
\resizebox{\hsize}{!}{
\includegraphics[angle=-90]{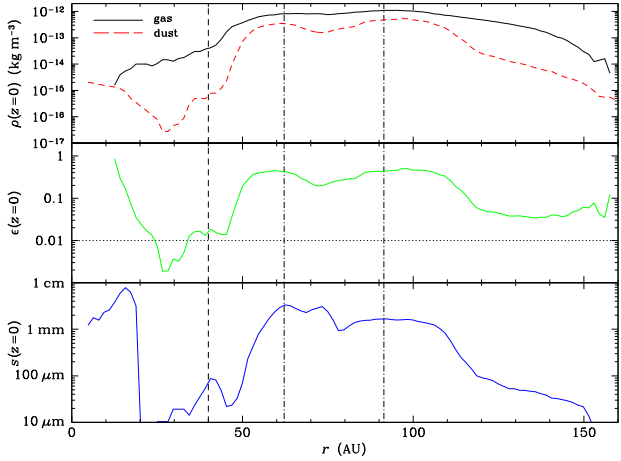}
}
\vspace{-2ex}
\caption{Azimuthally averaged radial profiles of the gas and dust volume densities (top), dust-to-gas ratio (middle), and mean grain size (bottom) in the midplane after 100\,000~yr. The vertical dashed line shows the planet orbital radius, the vertical dash-dotted lines show the locations of pressure maxima (see Fig.~\ref{Fig:P}) and the horizontal dotted line shows the initial dust-to-gas ratio.}
\label{Fig:rho}
\end{figure}

\begin{figure}
\centering
\resizebox{\hsize}{!}{
\includegraphics{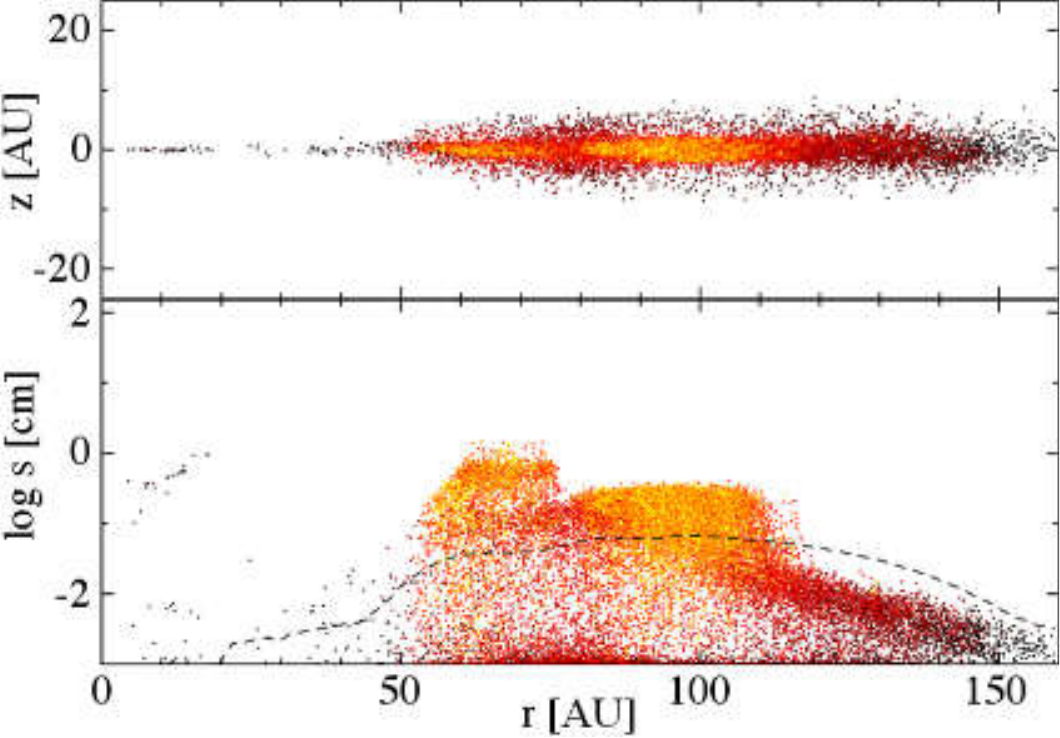}
}
\vspace{-4ex}
\caption{Meridian plane cut of the dust distribution (top) and radial grain size distribution (bottom) after 100 \,000 yr. The color represents the volume density and is the same as in Fig.~\ref{Fig:hydro}. The dashed line marks the grain size for which St = 1 in the disc midplane (see Sect.~\ref{sec:Discussion}).}
\label{Fig:SizeDist}
\end{figure}

Figure~\ref{Fig:hydro} shows the gas and dust volume densities $\rho_\mathrm{g}$ and $\rho_\mathrm{d}$, as well as the grain size $s$ in the disc midplane at the end of the simulation. The planet at 40~au has opened a gap in the gas phase and much of the disc interior to the gap has been accreted onto the star, leaving a central low-density area. In the dust phase, the gap is more prominent, with sharper edges. Some material remains in the inner disc and grains have accumulated at the dust trap located at the gap outer edge. These properties are commonly seen in studies of planet gaps in gas+dust discs \citep[see, e.g.,][]{Paardekooper2004,Fouchet2010}. An unexpected feature is the second high-density ring in the dust disc at $r\sim90$~au visible in the density map (centre panel of Fig.~\ref{Fig:hydro}) and in the radial profile of the midplane volume density (Fig.~\ref{Fig:rho}). Large grains (mm to cm in size) are abundant in the high-density regions (right panel of Fig.~\ref{Fig:hydro} and bottom panel of Fig.~\ref{Fig:rho}), where relative velocities between grains are lowest and grain growth is more efficient. The dust population is best characterised in Fig.~\ref{Fig:SizeDist}, showing the distribution of grains in the meridian plane and their radial size distribution. Two populations can be distinguished: grains that have been trapped in a narrow ring at the outer planet gap edge (at $r\sim60$~au) and have grown, and grains that have started to grow in the outer disc, migrated inwards, decoupled from the gas, and piled up in a wider ring $\sim$90~au from the star. We refer the reader to \citet{Gonzalez2015} for a detailed discussion of dust evolution, as well as Sect.~\ref{sec:Discussion}.

\subsection{Synthetic ALMA images}
\label{sec:Images}

We produced synthetic ALMA images for the final state of our disc. We first computed raw intensity maps from the resulting dust distribution using the 3D Monte Carlo continuum radiative transfer code \textsf{MCFOST} \citep{Pinte2006,Pinte2009}. It computes the scattering, absorption and re-emission of photons by dust grains assumed to be homogeneous spheres and composed of astronomical silicates, with optical properties derived according to Mie theory. These maps were then passed to the CASA 
simulator for ALMA to obtain synthetic images for a given observing configuration (wavelength, angular resolution, integration time). For more details about this procedure, see \citet{Gonzalez2012}.

\begin{figure}
\centering
\resizebox{\hsize}{!}{
\includegraphics{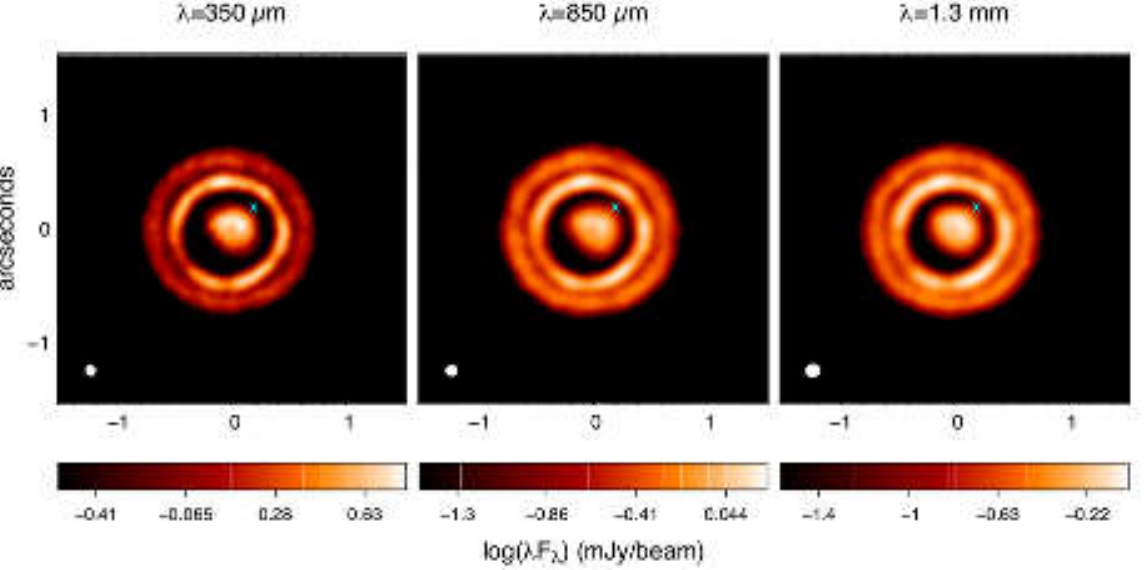}
}
\vspace{-4ex}
\caption{Simulated ALMA observations of a disc viewed nearly face-on at $d=140$~pc and $\delta=-23^\circ$ for an integration time of 1~h and angular resolution of $0.1''$. \textit{From left to right:} $\lambda=350\ \mu$m, 850~$\mu$m, and 1.3~mm. The scale on each image is in arcseconds, with the beam size represented at its bottom left corner, and the colourbar gives the flux in mJy/beam. The planet's position is marked with a cross.}
\label{Fig:ALMA}
\end{figure}

\begin{figure}
\centering
\resizebox{\hsize}{!}{
\includegraphics[angle=-90]{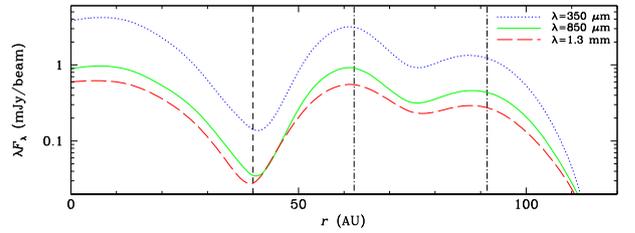}
}
\vspace{-2ex}
\caption{Radial cut of the brightness profile in the ALMA synthetic images for $\lambda=350\ \mu$m, 850~$\mu$m, and 1.3~mm. The vertical dashed line shows the planet orbital radius and the vertical dash-dotted lines show the locations of pressure maxima (see Fig.~\ref{Fig:P}).}
\label{Fig:ProfileVis}
\end{figure}

Figure~\ref{Fig:ALMA} shows the resulting synthetic images at three different wavelengths: $350~\mu$m, $850~\mu$m and 1.3~mm. They were computed for a nearly face-on orientation, a distance $d=140$~pc and a declination $\delta=-23^\circ$ (for which the source passes through the zenith at the ALMA site), an integration time $t=1$~h and angular resolution $\theta=0.1''$ \citep[found to be optimal observing parameters for gap detection by][]{Gonzalez2012}. We also plot in Fig.~\ref{Fig:ProfileVis} radial cuts of the brightness profile for the three wavelengths. All images show the central emission from the dust in the inner disc (interior to 20~au, see Fig.~\ref{Fig:hydro}), the deep and wide planet gap (centred on 40~au), a bright ring corresponding to the outer edge of the planet gap (near 60~au), a second shallower and narrower apparent gap (centred on $\sim75$~au, which happens to be near the 2:5 resonance with the planet), and finally a second fainter ring (centred on $\sim90$~au). The synthetic observations thus recover all features of the dust distribution. Note that a higher disc temperature would increase the computed fluxes, but would preserve the gap structures (see Sect.~\ref{sec:Hydro} for the influence of $T$ on the dynamics).

\section{Discussion}
\label{sec:Discussion}

\begin{figure}
\centering
\resizebox{\hsize}{!}{
\includegraphics[angle=-90]{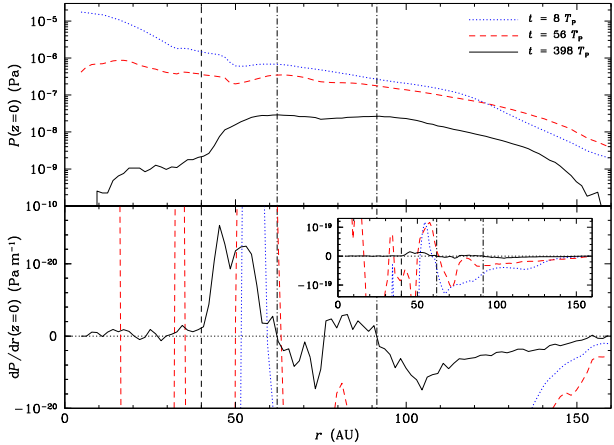}
}
\vspace{-2ex}
\caption{Azimuthally averaged profiles of the gas pressure (top) and pressure gradient (bottom) in the midplane after 8, 56 and 398 planet orbits ($\sim2\,000$, 14\,000 and 100\,000~yr). The vertical dashed line shows the planet orbital radius and the vertical dash-dotted lines show the locations of pressure maxima at the end of the simulation. The inset shows a larger vertical scale.}
\label{Fig:P}
\end{figure}

We find that, in a CTTS disc containing a 5~$M_\mathrm{J}$ planet on a fixed circular 40~au orbit, a fragmentation threshold of $\vfrag=15$~m\,s$^{-1}$ produces two almost axisymmetric gap and ring pairs in the corresponding synthetic images at (sub)millimetre wavelengths (Fig.~\ref{Fig:ALMA}). What is their origin? Lower continuum emission at a given location corresponds to lower dust opacity at millimetre wavelengths, which can be caused by a local minimum of the dust density or by grain sizes outside of the range contributing significantly at these wavelengths (i.e. 100~$\mu$m to a few mm, for the dust optical properties assumed here), or both. It is important to note that the two gaps in our disc have different origins.

The planet is massive enough to carve a gap in the gas phase and produce a local pressure maximum at the outer gap edge ($r\sim60$~au) very early in the disc evolution -- see the midplane pressure and pressure gradient profiles in Fig.~\ref{Fig:P}. Dust particles are trapped at the pressure maximum, accumulate, and grow more efficiently as they concentrate. This so-called particle trap generates the first ring just outside the planet gap \citep[see also][]{Pinilla2012gap}.

The second ring, on the other hand, is caused by a grain pile up rather than a trap due to the planet. Grains can grow when $\vrel$ (which scales as $c_\mathrm{s}\propto r^{-1/2}$ in our disc) is small enough to stay below $\vfrag$, i.e. in the disc outer regions. The growth timescale varies as $r^{3/2}$ \citep{Laibe2008}, it is thus shorter at the inner boundary of this growth region. Growing grains migrate inwards, more rapidly as their size approaches the value for which $\mathrm{St}=1$ (plotted as a dashed line in Fig.~\ref{Fig:SizeDist} at the end of the simulation, see \citealt{Gonzalez2015} for the time evolution). As grains continue to grow, they overcome the $\mathrm{St}=1$ regime, causing them to decouple from the gas, which slows their radial migration and results in a pile up near 90~au. This enhances the local concentration of grains, and the dust-to-gas ratio approaches unity (see the middle panel of Fig.~\ref{Fig:rho}), which i) stops grain migration in the accumulation region \citep{NSH1986} and ii) starts to modify the gas density profile due to the backreaction of dust on gas. Indeed, the drag force between gas and dust is symmetric and in a dust concentration, it is gas that is dragged by the dust. This effect can only be seen in self-consistent simulations of the simultaneous evolution of gas and dust taking backreaction into account. In other words, the complex interplay between growth and migration induces dust self-accumulation, which causes the gas to accumulate at the dust concentration and form a second gas pressure maximum around $r\sim90$~au at late times (see Fig.~\ref{Fig:P}). This effectively results in a second particle trap, causing the second ring. This mechanism, predicted theoretically by \citet{Laibe2014}, is independent of the presence of a planet, and was also seen in the very inner disc for the growth-only simulations in \citet{Gonzalez2015}. Here, we identify for the first time a self-induced dust pile up in simulations taking fragmentation into account.

Counter-intuitively, the value of the fragmentation threshold plays an essential role in the generation of this second ring. The dust pile up occurs near the location where $\mathrm{St=1}$ for the largest grains, for which $\vrel$ is also equal to $\vfrag$, i.e. where $c_\mathrm{s}\sqrt{\alpha_\mathrm{\scriptscriptstyle SS}}\sim\vfrag$. For a sound speed given by $c_\mathrm{s}=c_\mathrm{s,0}(r/r_0)^{-q/2}$, this location is approximated by $r_\mathrm{pu}/r_0\sim(\alpha_\mathrm{\scriptscriptstyle SS}\,c_\mathrm{s,0}^2/\vfrag^2)^{1/q}$. In our disc and for $\vfrag=15$~m\,s$^{-1}$, $r_\mathrm{pu}\sim90$~au. For larger values of $\vfrag$, grains would pile up at smaller orbital distances ($\sim50$ and $\sim30$~au for $\vfrag=20$ and 25~m\,s$^{-1}$) but they are trapped at the planet gap outer edge before they can decouple. There is only one ring in this case. On the other hand, for smaller $\vfrag$, only grains at very large orbital distances can grow, but because of their even lower growth rate they do not reach $\mathrm{St}=1$ and do not decouple before the end of the simulation \citep[see][for illustrations of both cases]{Gonzalez2015}. $\vfrag$ ranges from 1 and 10~m\,s$^{-1}$ for compact silicate and icy grains \citep{Blum2008} to several tens of m\,s$^{-1}$ for porous aggregates \citep{Yamamoto2014}. The value of 15~m\,s$^{-1}$, producing a second ring that is well separated from the outer gap edge, would match moderately porous ice. However, different values of $\vfrag$, or similarly of $\alpha_\mathrm{\scriptscriptstyle SS}$ or the disc temperature, can still lead to two separate rings if the planet location is changed.

The second apparent gap is therefore the region of lower dust density between both dust accumulations. Its properties differ from those of a planet gap: the decrease in the density of both gas and dust is modest and the grains have grown to moderate sizes. However, the observational signatures of both the planet gap and the apparent gap are very similar, both in the images (Fig.~\ref{Fig:ALMA}) and in the brightness profiles (Fig.~\ref{Fig:ProfileVis}), so that the shallower apparent gap seems to mimic a gap that would be carved by a lighter planet orbiting at $\sim75$~au (all the more so that it would lie close to the 2:5 resonance with the planet at 40~au).
Measuring this gap's width would give an estimate of this hypothetical planet's mass \citep{Crida2006}, but would not unambiguously prove its existence.
Cuello (in preparation) ran a series of simulations with two planets embedded in the same disc as ours. With planets of 5~$M_\mathrm{J}$ at 40~au and 1~$M_\mathrm{J}$ at 80~au, the outer planet gap has a similar depth in the gas as our apparent gap (also in agreement with \citealt{Kanagawa2015}) but a much deeper gap in the dust. Conversely, a lighter outer planet would be needed to reproduce the depth of our apparent gap in the dust, but it would carve an even shallower gap in the gas. If a suitable tracer of gas in the gap can be found, ALMA observations of both gas and dust and the comparison of gap depth in both phases would thus be able to distinguish an apparent gap from a planet gap.

\section{Conclusion}
\label{sec:Conclusion}

We have presented 3D hydrodynamical simulations of the evolution of gas and dust in a CTTS disc containing a planet which carves a gap in the disc. We self-consistently treat the interaction of gas and dust via aerodynamic drag and include grain growth and fragmentation. We then computed synthetic ALMA images from the resulting disc structure. We have shown that when the fragmentation threshold $\vfrag=15$~m\,s$^{-1}$, the dust dynamics results in the first documented self-induced pile up of grains when fragmentation is included, at large distances from the star. This forms a dense dust ring containing mm-sized grains in the outer disc, away from the planet gap and the dust trap at its outer edge. In the images, in addition to the gap+ring pair corresponding to the planet gap and its dense outer edge, this feature produces a second ring separated from the outer gap edge by an apparent gap, effectively resulting in a second gap+ring pair. This second, exterior gap does not contain a planet.

The recent spectacular ALMA image of the HL Tau disc \citep{HLTau2015} shows as many as seven gap and ring pairs. Is each of them an actual planet gap? The answer is probably no and we have shown here one example for which it is not the case. More gaps in discs are likely to be detected as ALMA is ramping up to full operation and they will have to be interpreted with care. Comparison of gap depths in both gas and dust would help to distinguish a planet gap from an apparent gap such as presented here.

\section*{Acknowledgements}

We thank the anonymous referee for a prompt and insightful report which helped us improve the discussion, in particular by suggesting the quantitative argumentation.
This research was partially supported by the Programme National de Physique Stellaire and the Programme National de Plan\'etologie of CNRS/INSU, France.
JFG thanks the LABEX Lyon Institute of Origins (ANR-10-LABX-0066) of the Universit\'e de Lyon for its financial support within the programme ``Investissements d'Avenir'' (ANR-11-IDEX-0007) of the French government operated by the ANR.
GL is grateful for funding from the European Research Council for the FP7 ERC advanced grant project ECOGAL.
STM acknowledges the support of the visiting professorship scheme from Universit\'e Claude Bernard Lyon 1.
Simulations presented in this work were run at the Service Commun de Calcul Intensif (SCCI) de l'Observatoire de Grenoble, France and at the Common Computing Facility (CCF) of LABEX LIO.
Figures~\ref{Fig:hydro} and \ref{Fig:SizeDist} were made with SPLASH \citep{Price2007}.



\bibliographystyle{mnras}
\bibliography{mmgaps}


\bsp	
\label{lastpage}
\end{document}